\documentclass[10pt,twocolumn,twoside]{IEEEtran}
\usepackage{subfigure}
\usepackage{epsfig}

\begin{document}

\setcounter{page}{144}
\title{Observation of Scattering and Absorption\\ Centers in Lead
  Fluoride Crystals}

\author{
 \thanks{Manuscript received August 9, 2000; revised October 31, 2000.}
 Patrick Achenbach \vspace{-5mm}
 \thanks{The author is with the Institut f\"ur Kernphysik, Johannes
  Gutenberg-Universit\"at, 55099 Mainz, Germany 
  (e-mail: patrick@kph.uni-mainz.de).}
 \thanks{Publisher Item Identifier S 0018-9499(01)02061-5.}
}
\pubid{0018--9499/01\$10.00~\copyright~2001 IEEE}
\markboth{IEEE Transactions On Nuclear Science, Vol. 48, No. 1,
  February 2001}{Achenbach: Scattering and Absorption Centers in Lead
  Fluoride Crystals}

\maketitle

\begin{abstract}
  For the first time, lead fluoride is used as a fast and compact
  material in electromagnetic calorimetry. Excellent optical and
  mechanical properties of the pure Cherenkov crystals are necessary
  for the A4 collaboration to perform a measurement of the nucleon's
  strange form factors. Visible scattering and absorption centers as
  well as surface damages have been investigated to characterize the
  quality of more than one thousand crystals. Besides, transmittance
  measurements have been performed on all crystals to reveal
  absorption bands produced by intrinsic or impurity related
  point-structure defects. As a consequence, 89 crystals had to be
  replaced by the Chinese manufacturer SICCAS.
\end{abstract}

\begin{keywords}Absorption bands, crystal defects, cubic lead fluoride, 
  electromagnetic calorimetry.
\end{keywords}

\section{Introduction}
\PARstart{L}{ead fluoride} (PbF$_2$) crystals have been studied as
detectors in nuclear physics for some time~\cite{Anderson90}.  Since
1993, large and transparent PbF$_2$ crystals have been available for
electromagnetic calorimetry. From that time on, several research
groups in high energy physics have taken into consideration to use
pure ionic or doped PbF$_2$ as a prospective calorimeter material and
beam tests with up to 25 crystals have been performed at different
laboratories~\cite{Woody93,Appuhn94}.  In 1999, the A4 collaboration
at the Mainz Microtron MAMI, Mainz, Germany began the construction of
the first PbF$_2$ calorimeter, which will eventually comprise 1022
crystals. In this parity violation experiment, polarized electrons are
scattered off a liquid hydrogen target to examine the possible
contribution of strangeness to the nucleon's Dirac and Pauli form
factors. The requirements on energy resolution, timing, compactness
and radiation resistance raised by the measurement of ppm asymmetries
are extremely high. For instance, an energy resolution $\sigma_E/E \le
3.5\% / \sqrt{E[\mathrm{GeV}]}$ in a $3 \times 3$ crystal matrix has
to be achieved to separate background events from elastic scattering
processes. Furthermore, the detector signals have to be shorter than
20\,ns otherwise the measured asymmetries might be falsified by pileup
losses. Only PbF$_2$ crystals of highest quality could fulfill those
demands.

The remainder of this paper is organized as follows: Section {\sc II}
reviews some basic properties of the PbF$_2$ compound and reports
technical aspects of the crystals' growing method. In Section {\sc
III}, scattering and absorption centers found in the crystals will be
described. Absorption bands and spatial zones of absorption have been
noticed in measurements of the longitudinal and transverse
transmittance. A brief summary of those findings is given in Section
{\sc IV}. Observed flaws will be documented in Section {\sc V}. The
last section provides the conclusions and their implications to the A4
experiment.

\pubidadjcol
\section{Properties of the Lead Fluoride Crystals}
The compound has outstanding properties concerning its application as
a Cherenkov detector, among them a high transmittance extending below
270\,nm, a density $\rho$ of 7.77\,g/cm$^3$, a radiation length $X_0$
of only 0.93\,cm and a Moli{\`e}re radius $R_{\mathrm{M}}$ of
2.2\,cm. The compound is usually found in a cubic lattice structure
($\beta$-PbF$_2$), but at high temperatures the orthorhombic
$\alpha$-phase appears. Since the raw materials are comparatively
cheap, the crystals could be produced at modest costs. The Shanghai
Institute of Ceramics and Chinese Academy of Sciences, Shanghai, China
(SICCAS) has grown more than 1100 $\beta$-PbF$_2$ mono-crystals under
contract with the Institut f\"ur Kernphysik, Mainz, Germany. Those
crystals have the geometry of pyramidal sectors with trapezoidal
basis. Their average cross-section is $25 \times 25$\,mm$^2$ at the
front face and $30\times30$\,mm$^2$ at the rear face and their lengths
vary between 150.0 and 185.4\,mm, corresponding to $16 -
20\,X_{\mathrm{0}}$.

The Bridgman method used by SICCAS allowed them to grow several
PbF$_2$ ingots at the same time. To avoid multiple appearances of
crystallization centers the prepared seeds are usually held at
temperatures below the melting point of 1097 K whereas the surrounding
melt is kept at higher temperatures. Using the Bridgman method the
growing ingots are fixed in their positions and the movable crucible
with the raw materials is pulled very slowly through a heating zone
which is controlled by radio-frequency heaters. Like the alternative
Czochralski method this technique has the disadvantage that the
crystals could be spoiled by reception of impurities from the
crucible. For this reason, SICCAS has investigated procedures to
reduce the number of impurities in the PbF$_2$ crystals to an
acceptable level. One solution is the application of post-growth
treatments where the crystals are annealed in a gaseous atmosphere,
which can remove oxide and hydroxide ions from the
crystals~\cite{SICCAS95}.  It is evident that the quality of the
crystals had to be controlled carefully after their delivery to
Mainz. Crystals had to be selected which met the stringent
requirements. A dedicated program of measurements included the
inspection of their longitudinal and transverse optical transmittance,
the determination of optical non-uniformities, the study of the
radiation resistance of small samples and the precise check of their
mechanical dimensions against the
specifications~\cite{Jahresbericht98}. Briefly, the radiation
resistance of the material was high, and induced damages could be
cured by optical bleaching with blue or near ultraviolet
light~\cite{Achenbach98}.  The crystals revealed an internal
transmittance of 100\%; defects like macroscopic cracks, texture, and
inclusions, however, have been observed in most of the them.

\section{Observed Scattering Centers}
The optical setup for studying the PbF$_2$ crystals consisted of a
{\em quartz tungsten halogen} lamp, a condenser close to the light
source, and an additional convex lens. The brightness was maximized
and a highly uniform illumination of the crystal fixed in the first
point of focus was provided. A screen was placed at the second point
of focus to obtain a magnified and sharp image of a cross section
through the crystal. In addition, photographs of illuminated crystals
were taken. This led to the observation of several types of scattering
centers and inclusions: needles, round shaped aggregates and
irregularly formed inclusions. Some of the inclusions were colorless,
some black, and only few white.

The needle-like inclusions started at the rear end of the crystals,
which was the seed side in the growing process, and extended some
centimeters. Their course was slightly bended, but followed the
longitudinal axis of the crystal.  Seemingly they originated from
one-dimensional lattice defects during an early stage of the crystal's
production. During the crystals' growth those defects have grown,
too. As an example distinctly illuminated needle-like inclusions can
be seen in Fig.\,1(a). Similar observations in PbF$_2$ crystals helped
the manufacturer SICCAS to identify the needle-like inclusions with
lead oxide or lead fluoro-oxide. \cite{SICCAS97}, presents an analysis
of {\em Energy Dispersive Spectra} measured at the {\em Laboratory of
Functional Inorganic Materials} and states that the inclusions are
mainly composed of lead, oxygen and fluor, and some of aluminum. A
research group from SICCAS made dedicated experiments in order to
reduce the forming of needle-like inclusions in PbF$_2$ crystals, even
if oxygen has been present in the melt. They succeeded by doping
certain amounts of deoxidants into the system, by increasing the
temperature gradient near the border between the growing mono-crystal
and the melt as well as by reducing the speed of the
crucible~\cite{SICCAS97}.  Nevertheless, about sixty of the crystals
delivered to Mainz showed needle-like inclusions. Fortunately, neither
did the inclusions cover the whole crystal volume, nor did they reduce
the transmittance significantly.

\begin{figure}[t]
  \begin{center}
    \subfigure[]{\epsfig{file=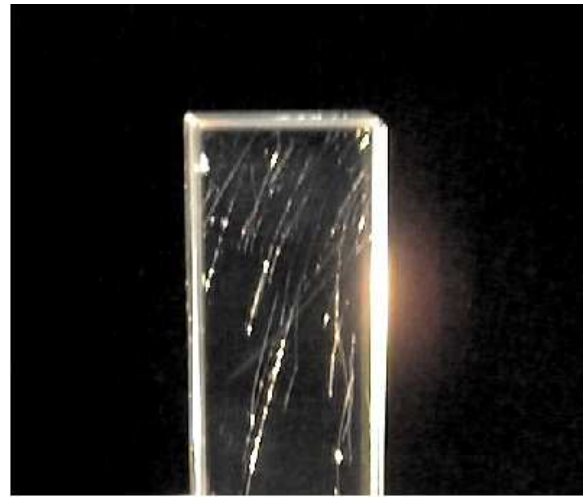, 
	width= .86\columnwidth}}\\ \vspace{-6mm} 
    \subfigure[]{\epsfig{file=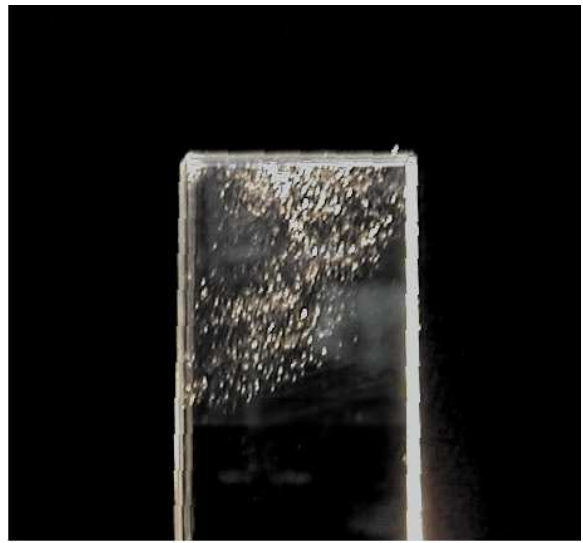, 
	width= .86\columnwidth}} \vspace{-4mm}
    \caption{Two examples of observed scattering centers.
    (a) Photograph of a crystal with needle-like inclusions.
    (b) Photograph of a crystal with round shaped aggregates.}
  \end{center} \vspace{-9mm}
\end{figure}

Another type of frequently observed scattering centers were round
shaped aggregates. One example of the 33 crystals where this type of
inclusion has been found is shown in Fig.~1(b). About 150 crystals
revealed point-like scattering centers, which were too small to reduce
the transmittance considerably. Fig.~2(a) presents a cross-section
through a crystal with some black inclusions at the left corner and
other inclusions at the bottom. In total, five crystals had black,
irregularly shaped inclusions located directly below their
surface. These defects supposedly originated in impurities that were
present in the crucible during the growth process. In \cite{SICCAS97},
some irregularly shaped inclusions are identified with orthorhombic
PbF$_2$ aggregates. This may result from an incomplete phase
transition of $\alpha$-PbF$_2$ into the $\beta$-phase during the
heating of the raw material.

Apart from other defects, Fig.~3(a) shows larger aggregates grouped in
several spatial domains of the crystal. Four crystals exhibited a
texture of small microcracks that spread over a volume of some cubic
centimeters. A further, but rare, observation was the appearance of
white inclusions measuring about $1-2$\,mm. This was noticed at the
rear faces of some crystals after having coupled them to a
photomultiplier, one is shown in Fig.~2(b). Presumably, the coupling
procedure led to the transformation.

\section{Observed Absorption Centers}
\begin{figure}[t]
  \begin{center}
    \subfigure[]{\epsfig{file=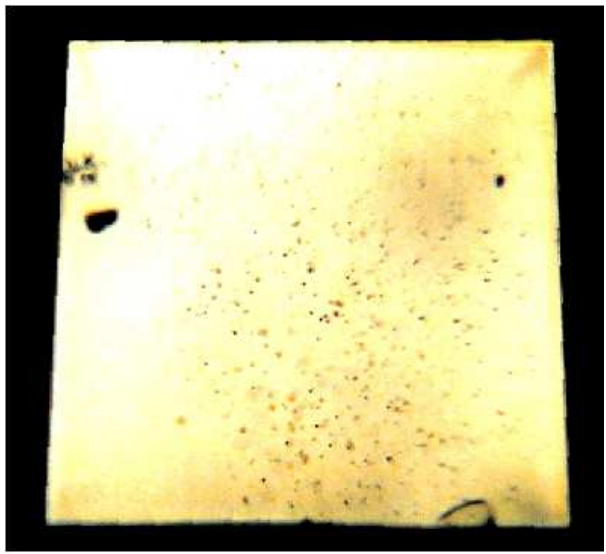, 
	width= 0.9\columnwidth}}\\ \vspace{-6mm}
    \subfigure[]{\epsfig{file=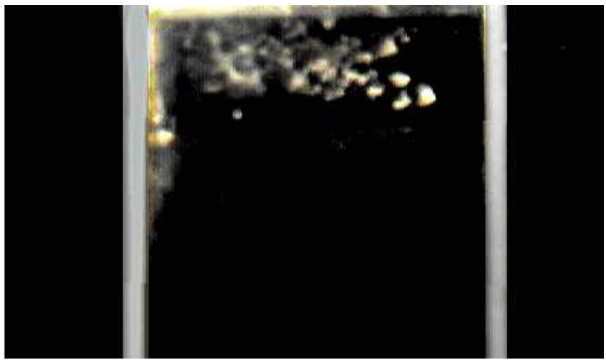,
	width= 0.9\columnwidth}} \vspace{-4mm}
    \caption{Two examples of rarely observed scattering centers.
    (a) Photograph of a projected cross section through a crystal with
	different inclusions.
    (b) Photograph of a crystal with white inclusions at its rear 
	face (seen as reflection).}
  \end{center}
  \vspace{-8mm}
\end{figure}

In a real crystal, the symmetry of the lattice is always broken at
some points. Those defects can lead to a reduction of the optical
transmittance if they appear in high concentrations. Point-structure
defects could be divided into impurity related defects, which are
localized substitutional or interstitial, and intrinsic defects like
trapped holes or interstitial Pb$^{2+}$ or F$^-$ ions. The theoretical
understanding of {\em localized impurities} in crystalline lattices is
a difficult task and details could be determined only by calculating
the embedding potential acting on the specific impurity. {\em
Intrinsic defects} and electron density distributions could be modeled
more accurately by quantum chemistry computer codes and it is known
that the concentration of intrinsic defects, which are sometimes
called thermal defects, increases exponentially with the
temperature. A good measure of the defect concentration is the
intensity of the related absorption bands in the transmittance
spectra. The band characteristics depend very much on the quality of
the raw material and the conditions during the growth process.
Certain rare earth elements such as Cerium, Praseodymium, or Thulium
are known to cause absorption bands at specific wavelengths. Key
elements such as oxygen in the O$^{2-}$ form and hydrogen are
particularly important.

\begin{figure}[t]
  \begin{center}
    \subfigure[]{\epsfig{file=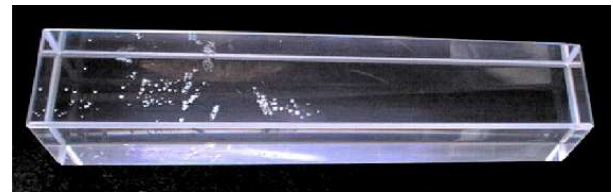, 
	width= 0.9 \columnwidth}}\\ \vspace{-6mm}
    \subfigure[]{\epsfig{file=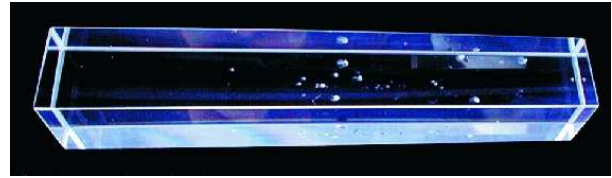,	
	width= 0.9 \columnwidth}} \vspace{-4mm}
    \caption{Two further examples of observed scattering centers and
    flaws. 
    (a) Photograph of a crystal with round shaped aggregates occuring 
	in groups.
    (b) Photograph of a crystal with surface damages and
    	flaws.}
  \end{center}
  \vspace{-6mm}
\end{figure}

The transmittancies of all PbF$_2$ crystals have been determined
between 200 and 800\,nm by using a commercial double beam
spectrophotometer Shimadzu UV-2101 PC.  A very large sized compartment
with a maximum probe length of 300\,mm was used to measure the
transmittance, not only transverse, but also longitudinal to the
crystals' axes. An integrating sphere collected all light passing
through the samples to avoid corrections normally required by the
nonparallel sides of the crystals. Assuming that the crystals have two
parallel end faces, the reflections of the light beam limit the
maximum transmittance to about 82\% at 500\,nm. The average values of
the longitudinal transmittancies of all inspected crystals, including
the rejected ones, were determined to be 80.4\% at 800\,nm, 74.1\% at
400\,nm, 64.8\% at 330\,nm, and 15.1\% at 270\,nm. These results were
far better than the required specification limits of 75\%, 70\%, 58\%,
and 2\% at the specified wavelengths demonstrating the high optical
quality of a majority of the PbF$_2$ mono-crystals. Nevertheless,
about 50 crystals exhibited prominent absorption bands. Mostly they
were located at wavelengths of 295\,nm, 395\,nm and 460\,nm
corresponding to transition energies of 4.2\,eV, 3.1\,eV and 2.7\,eV,
respectively. In some cases broad but weak absorption bands were
present at 560\,nm (2.2\,eV) and 595\,nm (2.1\,eV). Fig.~4 shows some
selected transmittancies where absorption bands are
visible. Unfortunately, it was not possible to correlate specific
impurities or defects with the observed absorption bands.

The crystals were scanned transverse to the longitudinal axis in steps
of 5\,mm between the front and the rear side in order to investigate
spatial zones of absorption as well. A large fraction of the crystals
did not show any, but in a considerable number of the investigated
crystals spatial zones of absorption were found at different distances
to the crystals' seed sides. In Fig.~5 the transverse transmittance at
1, 4, 8 and 12\,cm distance is shown for four selected crystals. As an
example of a high spatial dependence of absorption centers crystal
S1~021 is shown in Fig.~5(d). Fig.~5(b) exhibits the transmittance of
crystal S3~058 with a small, but constant spatial gradient of
absorption centers from one side to the other. In contrast, the
measurements of crystal S6~063 reveal in Fig.~5(a) a strongly varying
spatial gradient of the transmittance. In crystal S7~110, an
absorption zone was found in the front part, but a high and uniform
transmittance at larger distances as indicated by the three curves in
Fig.~5(c) which are laying on top of each other. Such an absorption
zone that is located at that part of the crystal, which was grown in a
later stage of the production process, was typical for the majority of
the affected crystals. The measured transmittancies of the crystals in
Fig.~4 showed this type of spatial dependence as well.

\begin{figure}[t]
  \onecolumn
  \begin{center}
    \subfigure[]{\epsfig{figure=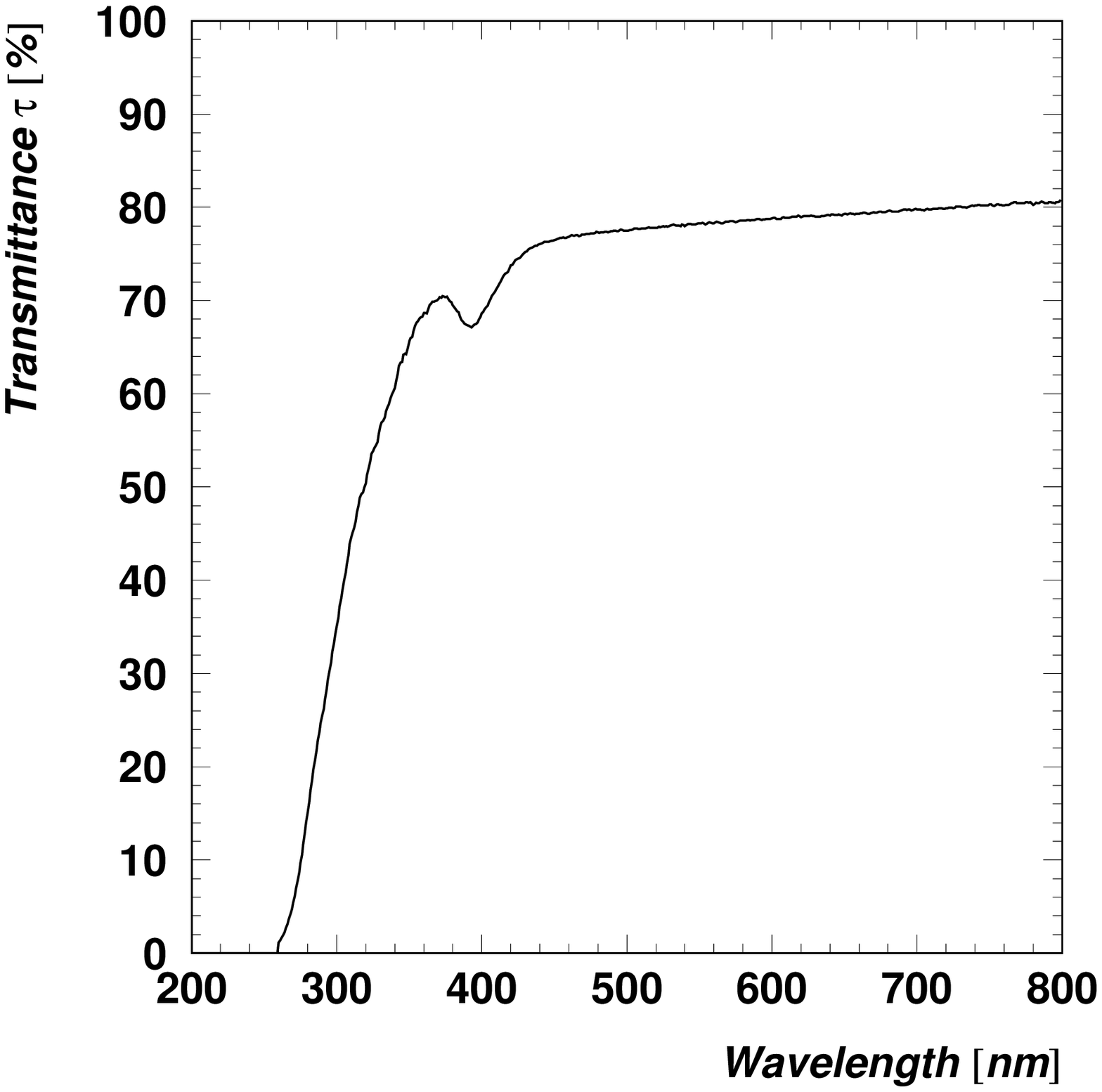, 
	width= 0.4 \textwidth}}
    \subfigure[]{\epsfig{figure=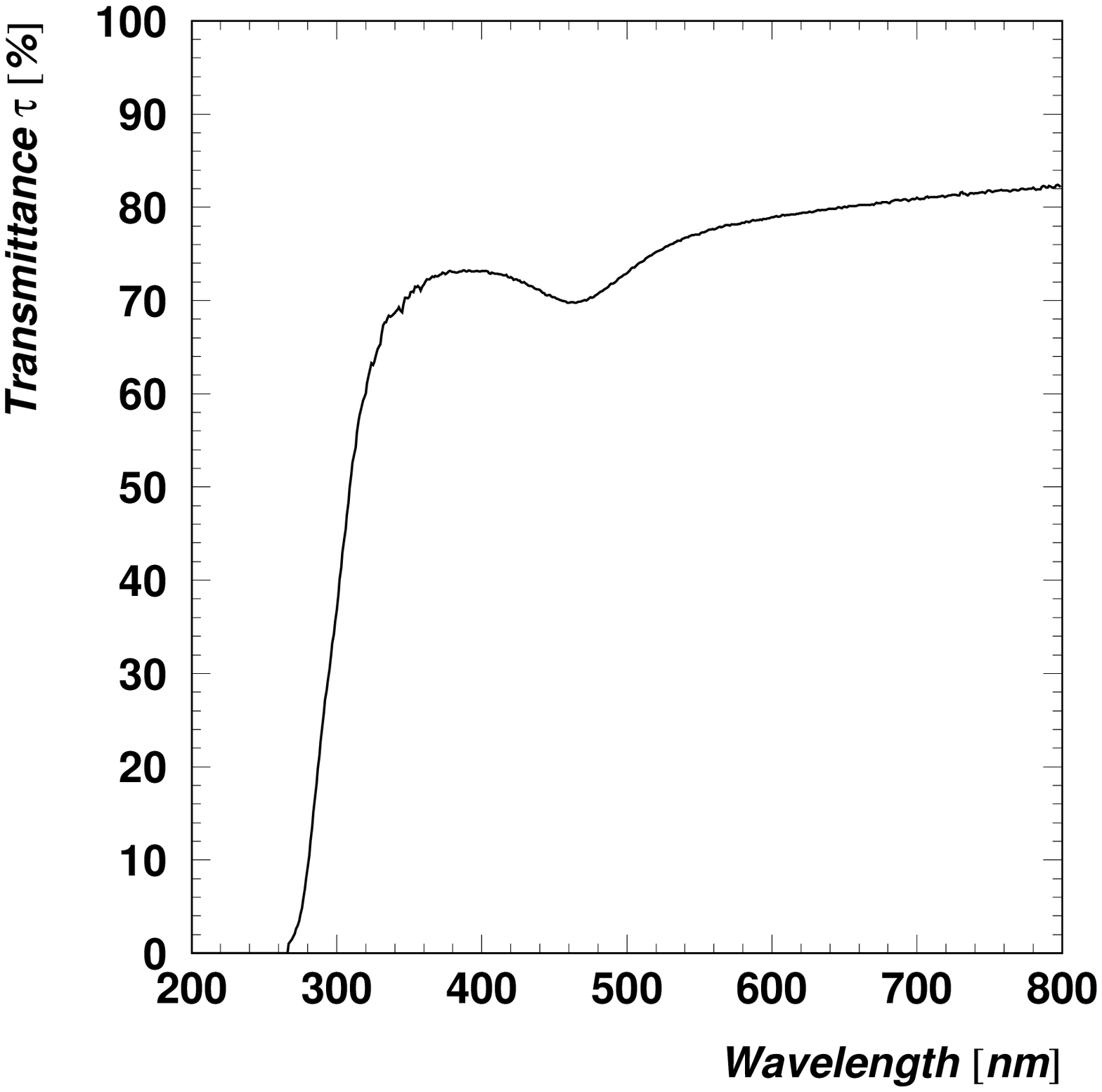, 
	width= 0.4 \textwidth}}\\ \vspace{-6mm}
    \subfigure[]{\epsfig{figure=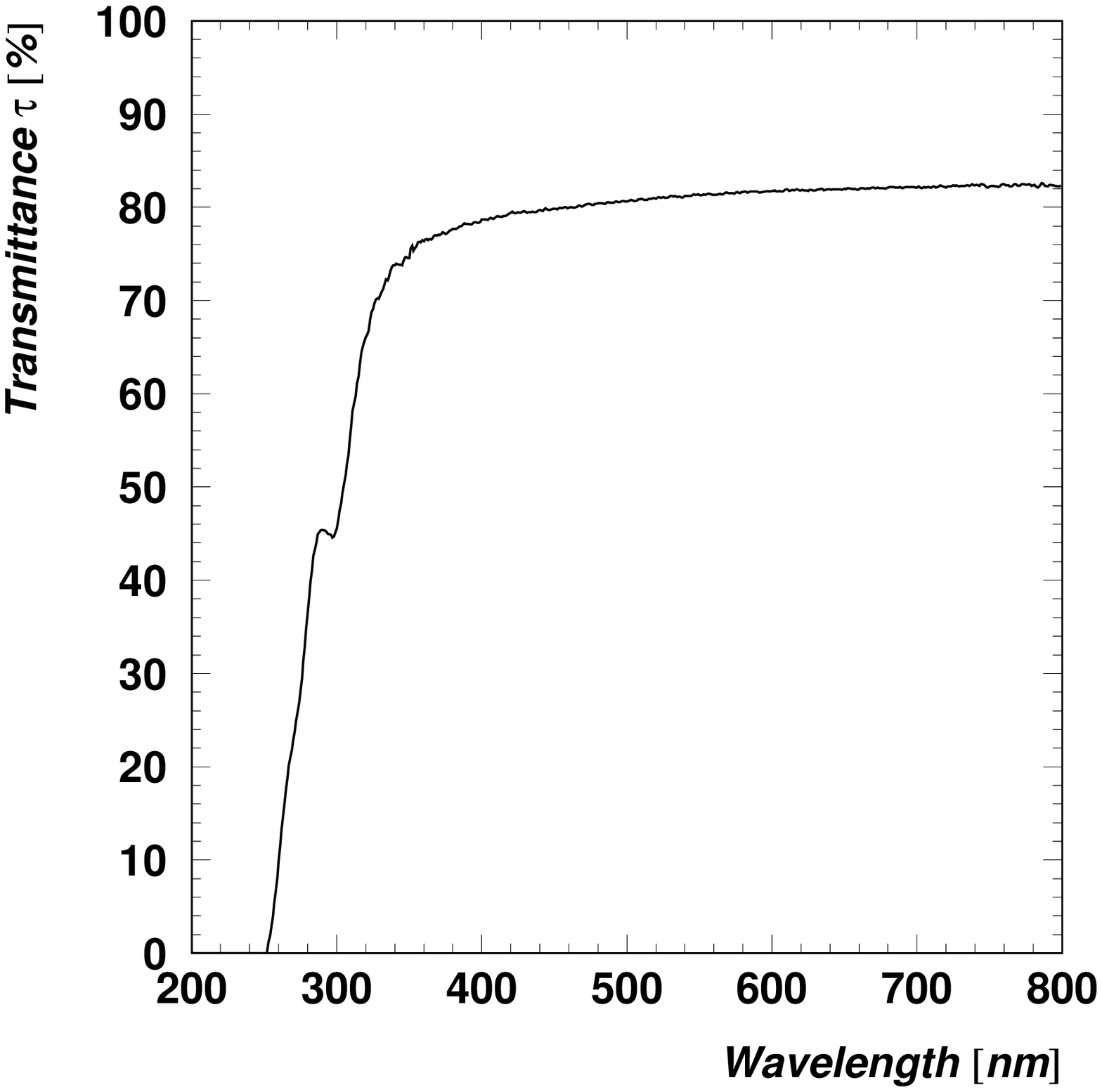, 
	width= 0.4 \textwidth}}
    \subfigure[]{\epsfig{figure=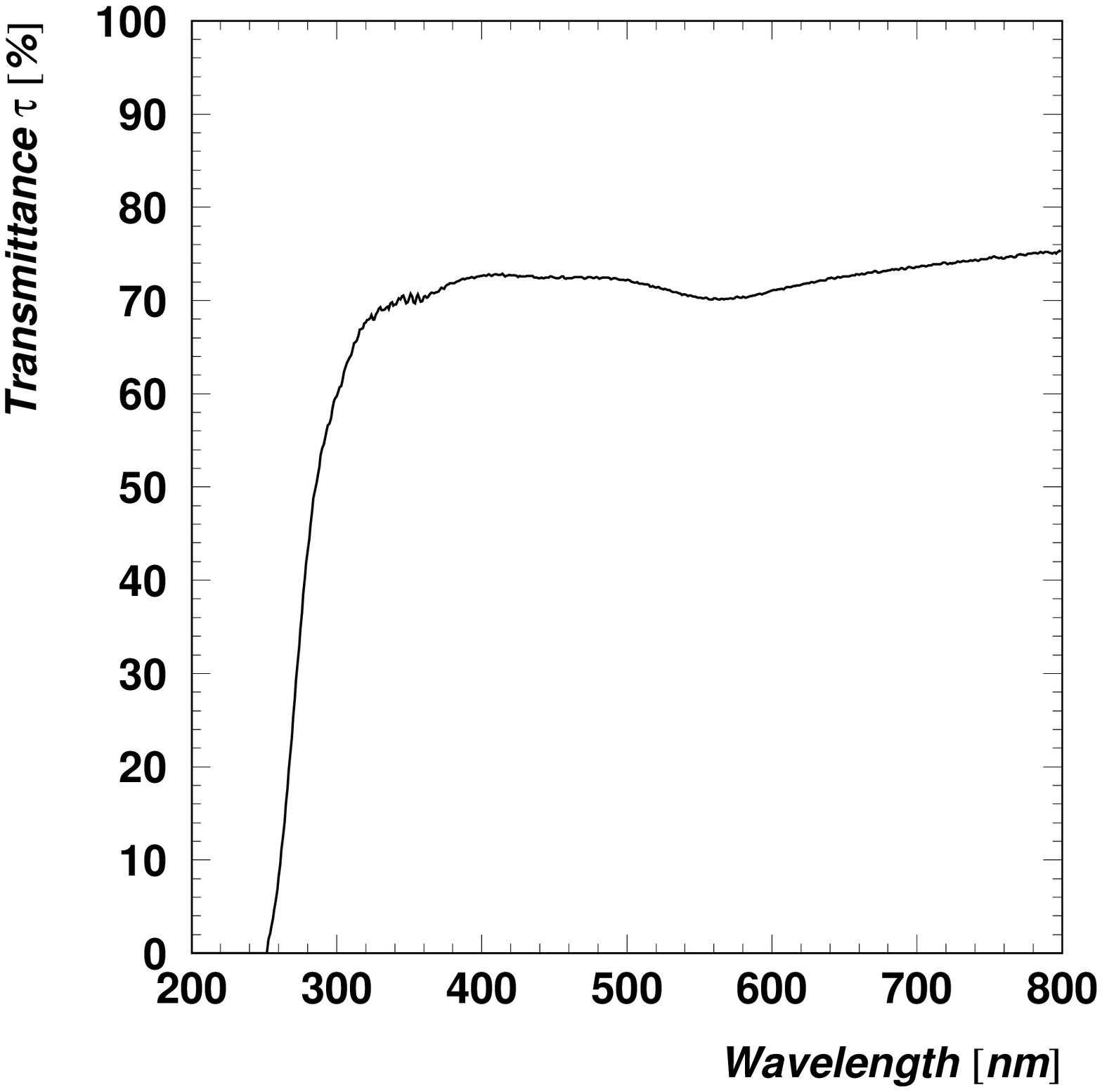, 
	width= 0.4 \textwidth}} \vspace{-4mm}
    \caption{The transmittance measured along the longitudinal axis of
    	four selected crystals with prominent absorption bands at
    	specific wavelengths.
    (a) Crystal S1 036.
    (b) Crystal S5 006.
    (c) Crystal S4 031.
    (d) Crystal S1 106.}
  \end{center}
  \twocolumn \vspace{-8mm}
\end{figure}

\vfill

\section{Observed Crystal Flaws}
Even flaws which cannot be explained by impurities or lattice defects
might affect the quality of the crystals. Stress regions inside the
bulk material which indicate domains of possible micro or macro cracks
could be made perceptible by looking for birefringence in the
crystals. Birefringence of light beams appears when the isotropic
lattice of a crystal is deformed by external or internal
forces. Putting two perpendicular polarizers in front of and behind
the crystal in the described optical setup exhibited colorful domains
at a small number of crystals, where the shape and appearance of the
domains were linked to the internal stress. 

\newpage \mbox{} \vfill

In Fig.~6(a), stress regions in the corners of a crystal can
be recognized. They were characteristic for the greatest part of the
investigated crystals and are supposedly due to the mechanical
treatments like cutting or polishing. Fig.~6(b) presents a large crack
diagonal to the longitudinal axis of the crystal which has probably
been caused by internal stress. Fig.~6(c) shows an externally stressed
crystal which was exposed to an electron beam of 20\,nA intensity,
causing strong temperature gradients by the high amount of deposited
energy in a short time.

Smaller flaws or cracks in the material could be found in almost 300
crystals. Again, thermal stress due to high temperature gradients,
which can occur during the annealing process, can be considered as the
main reason for cracks in the crystals. Bigger splintering, as it is
seen in the crystal of Fig.~3(b), could hardly be found, whereas
smaller damages of the crystals' faces were stated very often. Visible
scratches, for instance, were observed at more than 200
crystals. Milky

\newpage

\noindent
streaks on the surface have been noticed at about 270 crystals, where
the transmittance transverse to the crystal axis degraded by more than
15\%. A typical crystal with milky streaks revealed transmittance
values of 60\% at 800\,nm, 50\% at 400\,nm and 40\% at 330\,nm. The
streaks appeared on the surface of the crystals and were caused by
humidity, since PbF$_2$ is slightly hygroscopic. They could not
dissolved with acetone or any other appropriate method. The crystals
most affected exhibited transmittancies below 30\% in the entire
spectrum.

\begin{figure}[t]
  \onecolumn
  \begin{center}
    \subfigure[]{\epsfig{figure=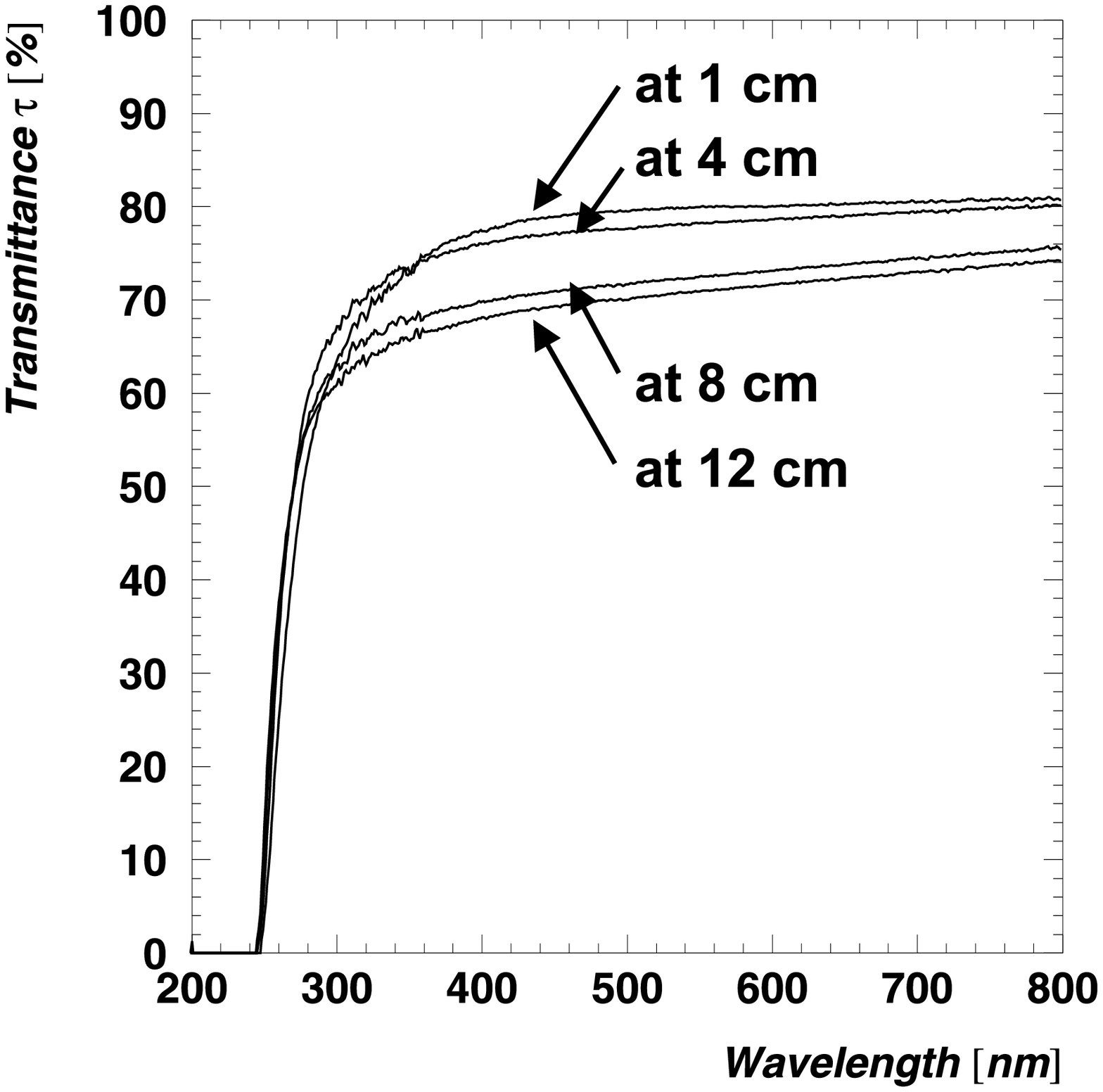, 
	width= 0.35 \textwidth}}
    \subfigure[]{\epsfig{figure=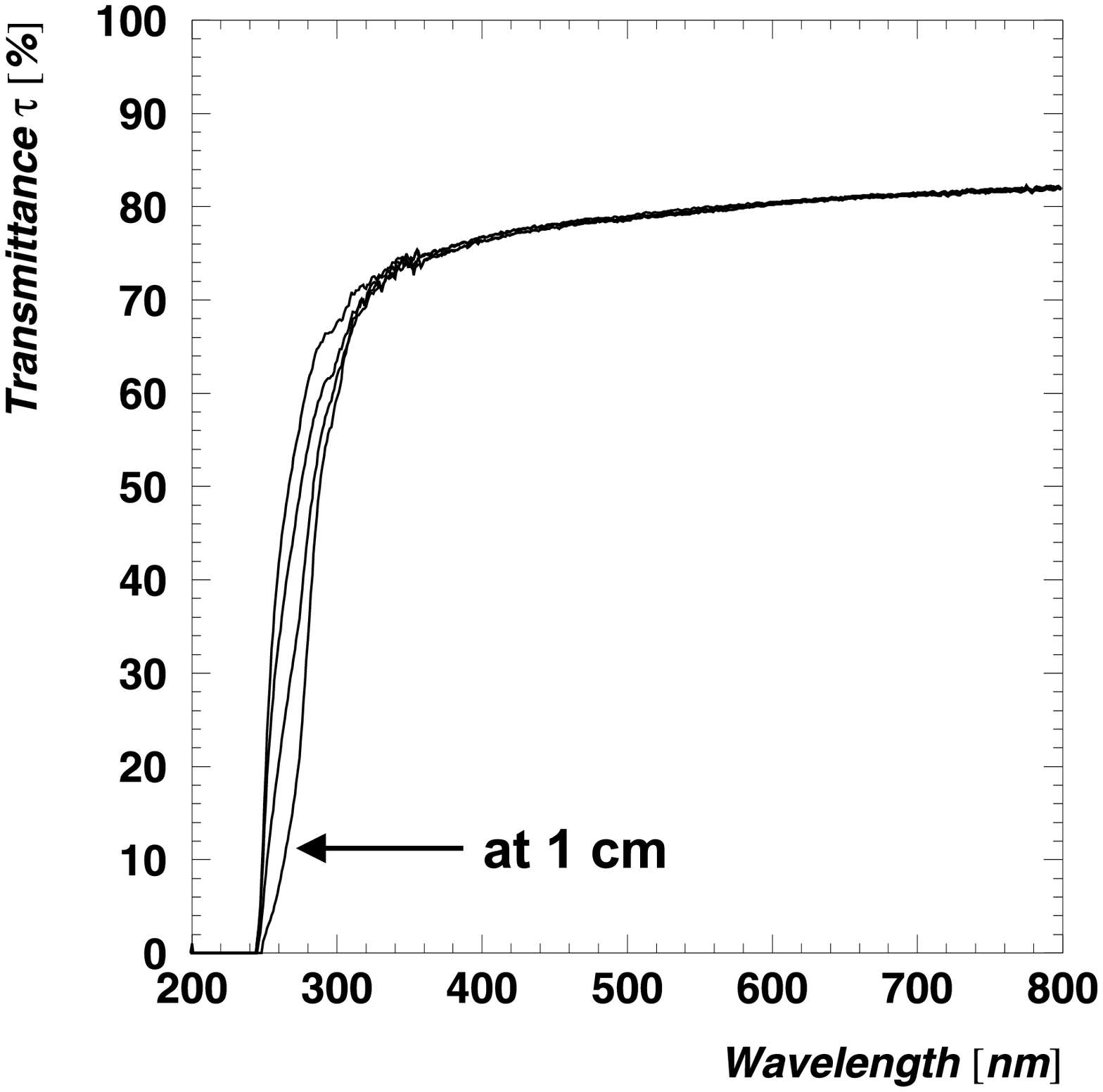, 
	width= 0.35 \textwidth}}\\ \vspace{-6mm}
    \subfigure[]{\epsfig{figure=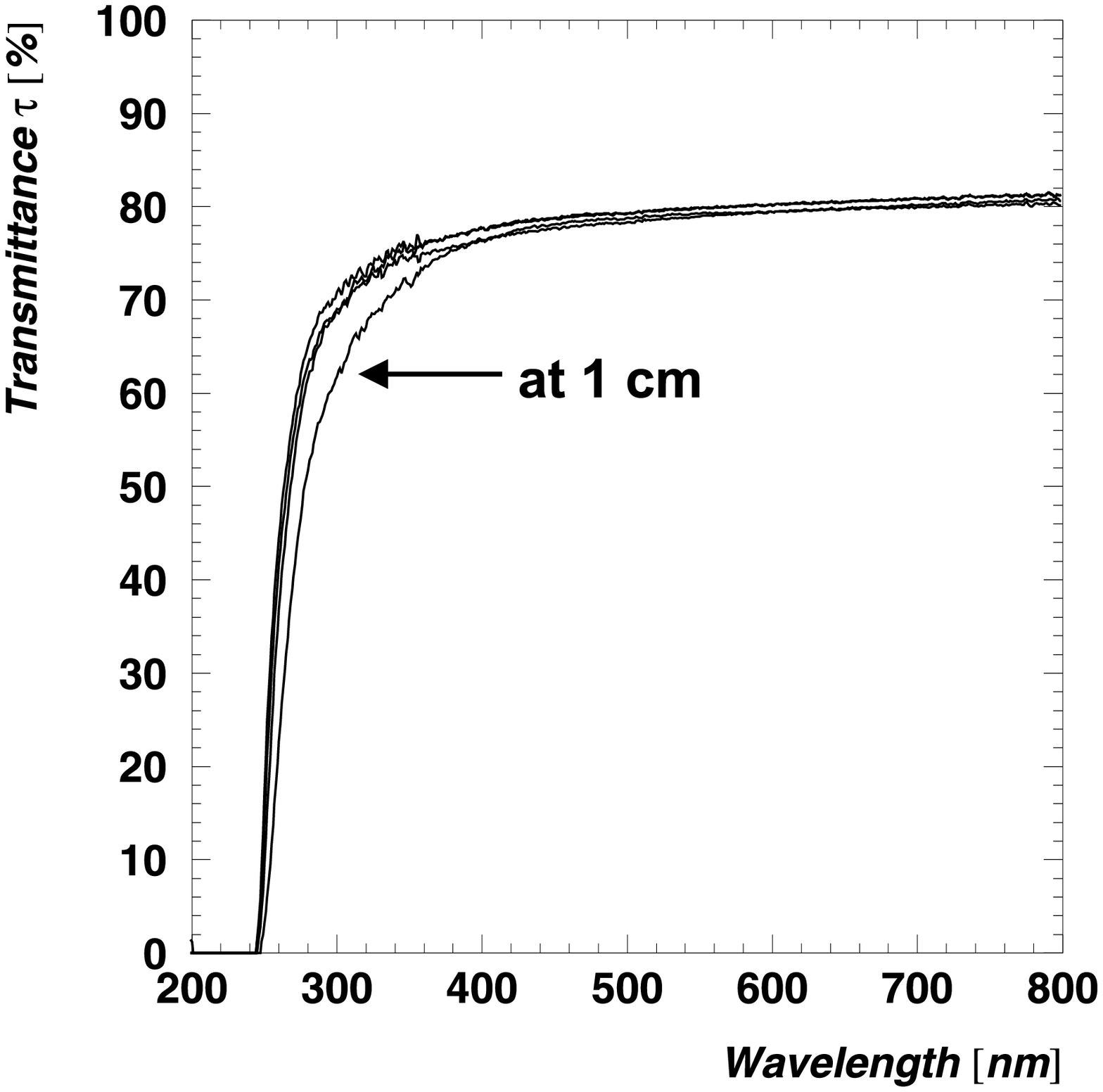, 
	width= 0.35 \textwidth}}
    \subfigure[]{\epsfig{figure=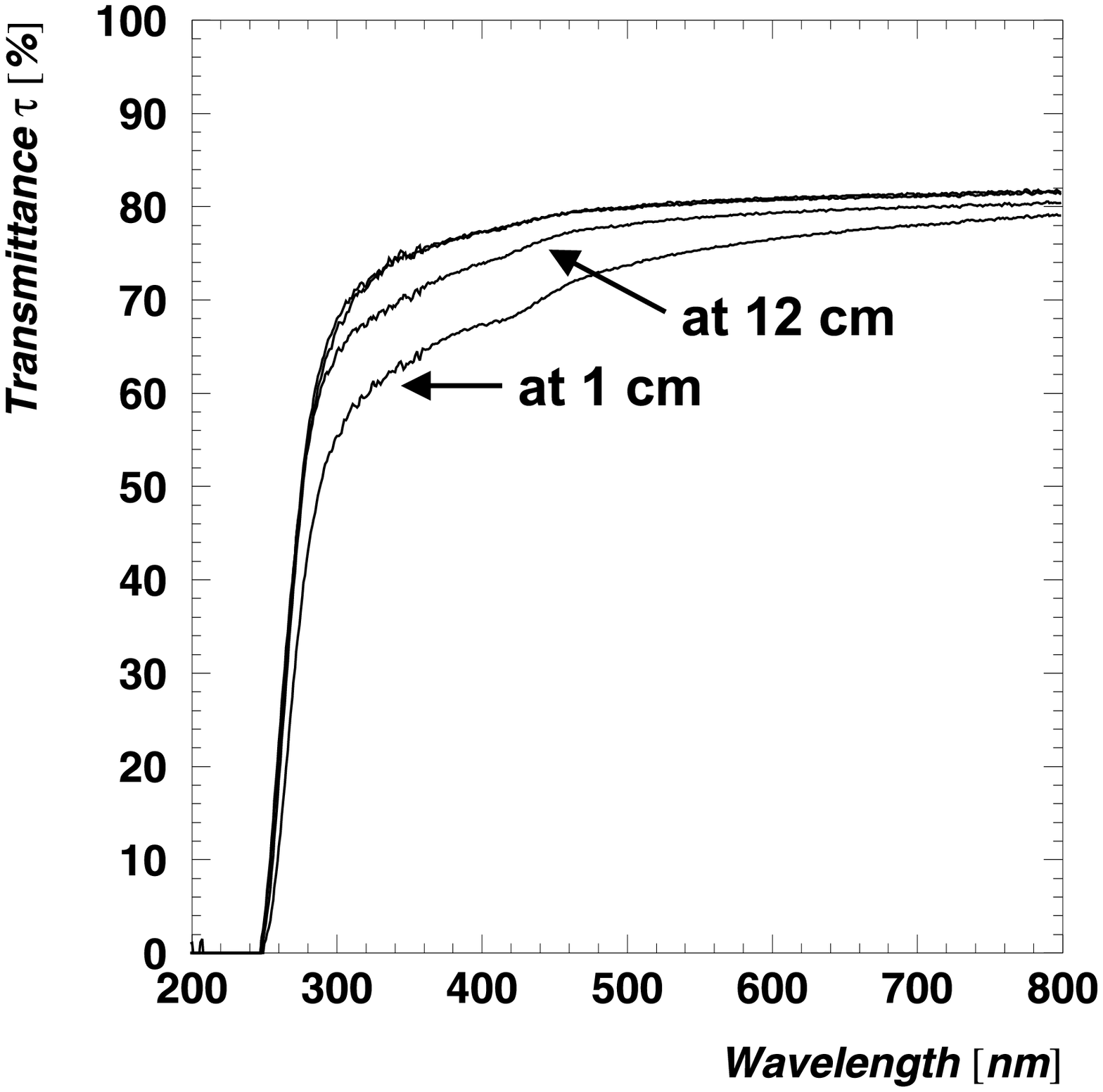, 
	width= 0.35 \textwidth}} \vspace{-4mm}
    \caption{The transmittancies measured transverse to the
    	longitudinal axis of four selected crystal at 1, 4, 8 and
    	12\,cm distance to the crystals' front faces.
    (a) Crystal S6 063.
    (b) Crystal S3 058.
    (c) Crystal S7 110.
    (d) Crystal S1 021.}
  \end{center}
  \twocolumn \vspace{-8mm}
\end{figure}

\vfill

\section{Conclusions}
The optical properties of 1100 PbF$_2$ crystals grown by SICCAS for
the A4 experiment have been studied. Minor defects like small
scattering centers or surface damages have been found in 87\% of the
crystals. In general, the optical transmittance of all the crystals
was high and close to the theoretical limit with a sharp band edge and
without absorption bands. A high concentration of intrinsic or
impurity related defects or scattering centers were found to be
responsible for

\newpage

\mbox{} \vfill

\noindent a low optical quality of a small number of crystals. In some
cases scattering centers even covered large parts of the
crystals. Additional measurements of the transmittance exhibited
absorption bands at specific wavelengths. Only about 140 crystals were
transparent without any defects. Since the Chinese manufacturer SICCAS
had guaranteed a certain optical quality, the replaced 89 crystals
which had revealed low transmittance curves, strong absorption bands,
or surface damages. The presented observations will allow the producer
to understand the impurity and defect structure of PbF$_2$
mono-crystals, and to improve their growing methods.

The measurements of the transmittance and the characterization of all
crystals allowed the A4 collaboration to select appropriate crystals
which met the stringent demands of its experiment. Combined with the
inspections of the mechanical dimensions using a computer-controlled
distance-measuring~\cite{Jahresbericht98} and the results of the
radiation resistance measurements~\cite{Achenbach98}, 171 crystals
were returned to SICCAS for reuse as raw crystal growing
material. After having finished this quality control program, the
setup of one-half of the calorimeter solid angle has been
completed. In July 2000, a successful data-taking

\newpage 

\mbox{}
\begin{figure}[t]
  \begin{center}
    \subfigure[]{\epsfig{ file=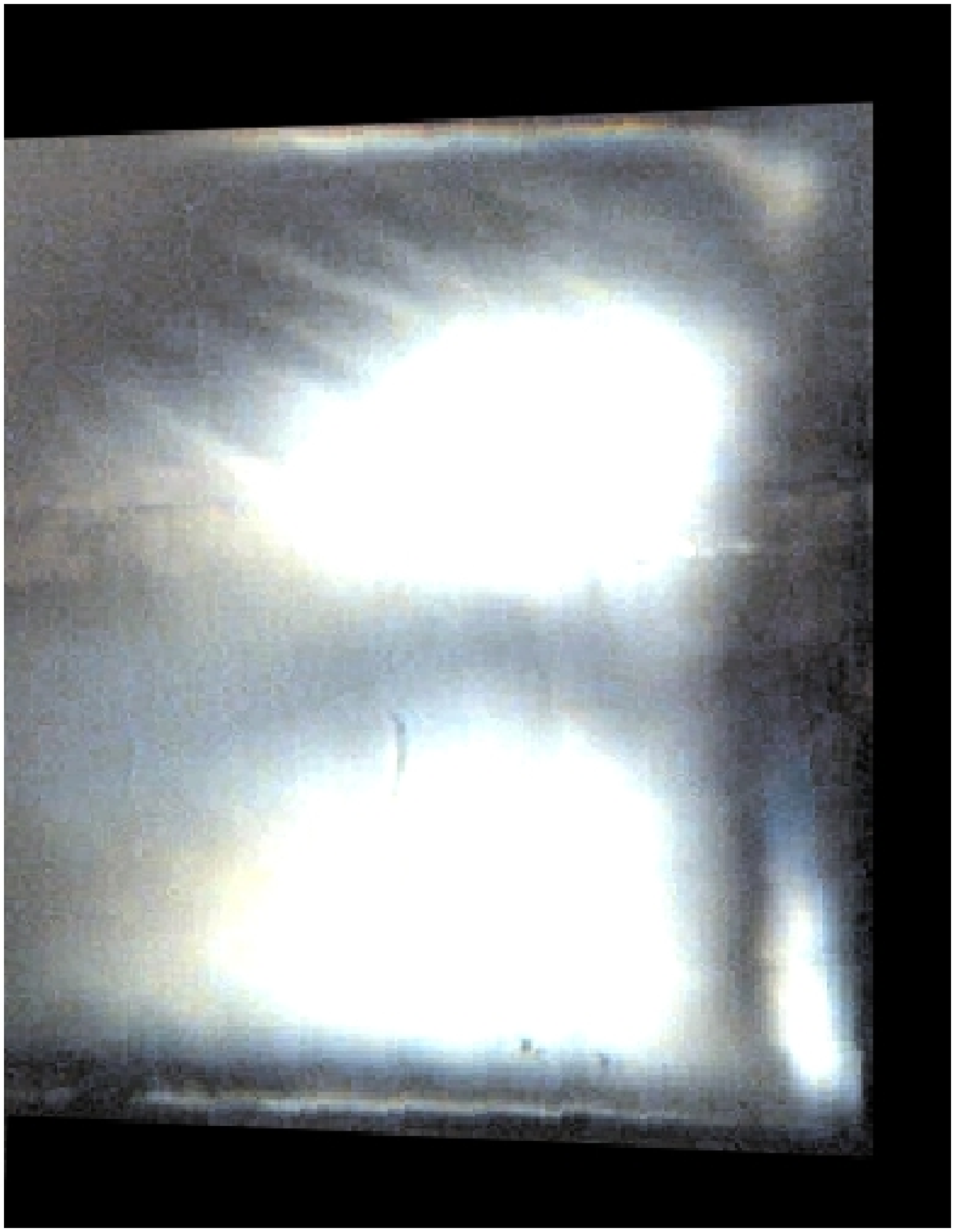, 
	height= 0.5 \columnwidth}}
    \subfigure[]{\epsfig{ file=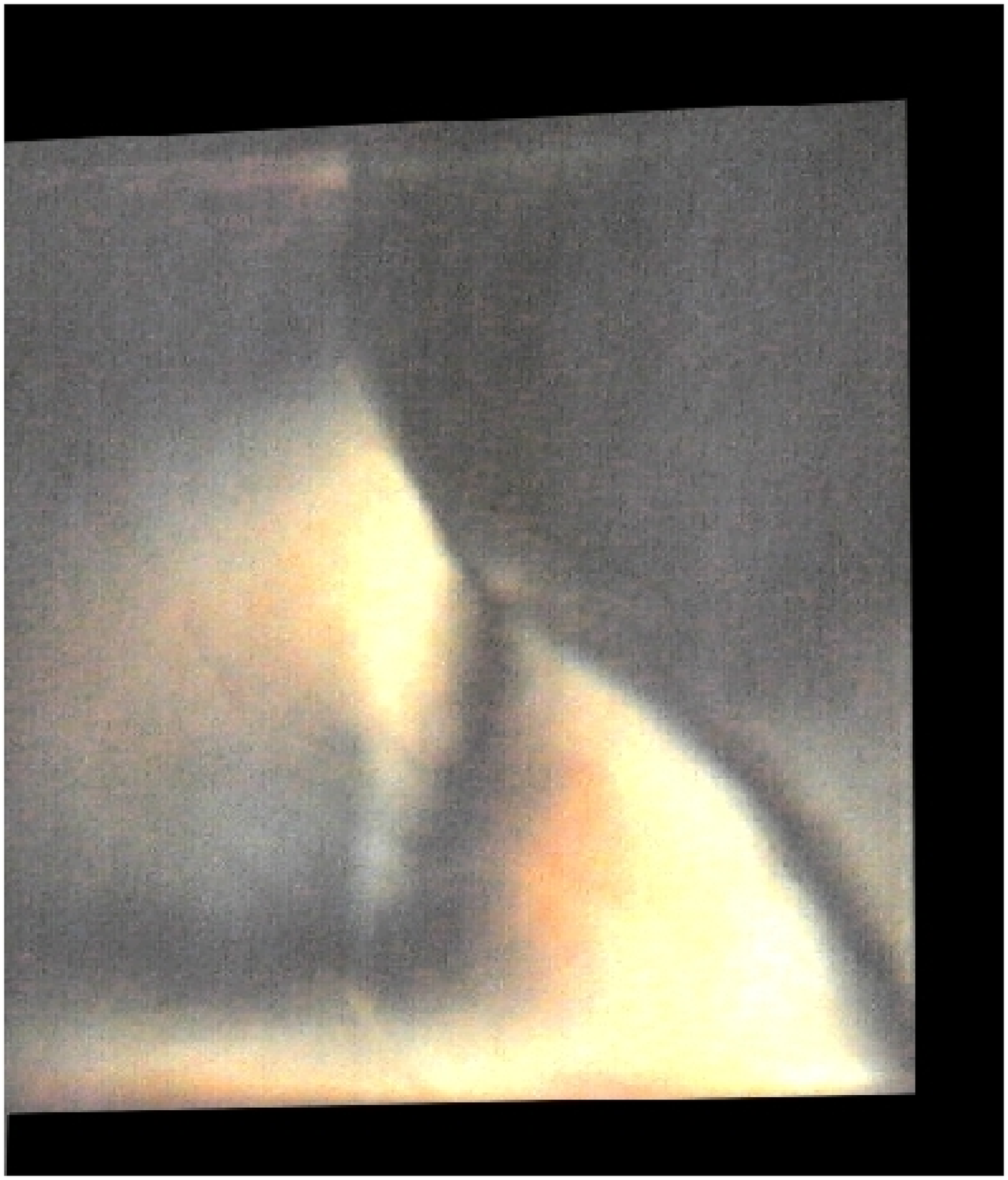, 
	height= 0.5 \columnwidth}}\\ \vspace{-6mm}
    \subfigure[]{\epsfig{ file=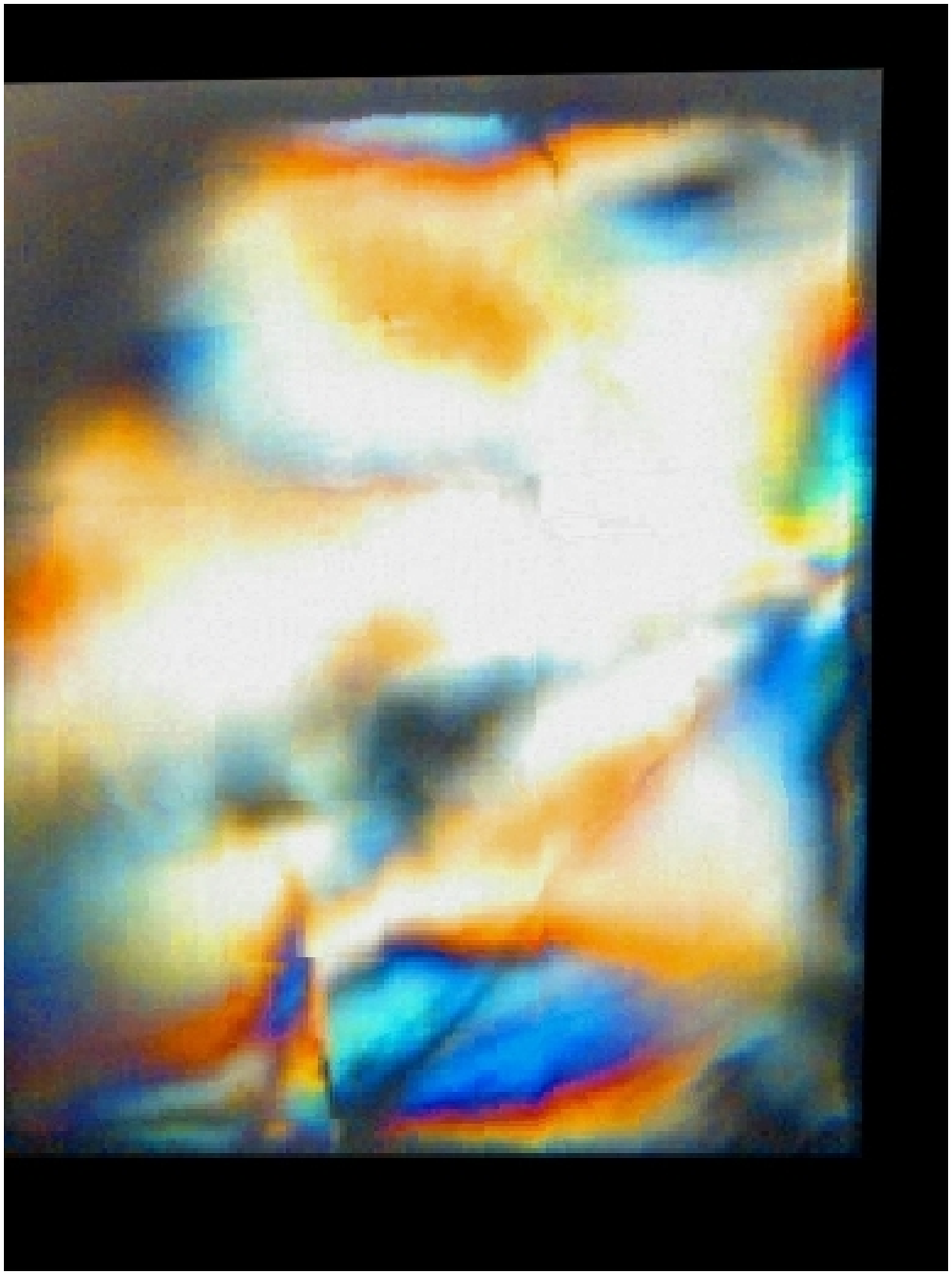, 
	width= 0.4 \columnwidth}} \vspace{-4mm}
    \caption{Flaws inside the crystals' bulk material have been
        observed in a small number of crystals. Two perpendicular
        polarizers allowed to visualize stress regions in the
        crystals.
    (a) Typical internal stress regions.
    (b) A diagonal crack caused by internal stress.
    (c) Texture of cracks and strong internal stress.}
  \end{center} \vspace{-4mm}
\end{figure}

\newpage

\noindent period with 511 PbF$_2$ detectors confirmed the excellent
optical properties of the selected crystals. Most of them fulfilled
the stringent requirement on energy resolution. The experiment is in
progress, and the collaboration will present results in a forthcoming
publication.

\vspace{9mm}


\end{document}